\documentclass[aps,preprintnumbers,amsmath,amssymb,nofootinbib,notitlepage,onecolumn]{revtex4-1}

\usepackage{amssymb}
\usepackage{graphicx}
\usepackage{amsmath}
\usepackage{multirow}
\usepackage{verbatim}
\usepackage[usenames,dvipsnames]{color}

\newcommand{\ion}[2]{{\text{#1{\sc #2}}}}

\newcommand{\nc}{\newcommand}

\nc{\ba}{\begin{eqnarray}}
\nc{\ea}{\end{eqnarray}}
\nc{\bc}{\begin{center}}
\nc{\ec}{\end{center}}
\nc{\ie}{\textit{i.e.}}
\nc{\del}{\partial}
\nc{\ud}{\mathrm{d}}
\nc{\mx}{m_\chi}
\nc{\feff}{f_{\mathrm{eff}}}
\nc{\geff}{g_{\mathrm{eff}}}
\nc{\zeff}{z_{\mathrm{eff}}}
\nc{\zmin}{z_{\mathrm{min}}}
\nc{\zmax}{z_{\mathrm{max}}}
\nc{\sv}{\langle \sigma v \rangle}
\nc{\pann}{p_{\rm ann}}

\begin{document}

\preprint{IFIC/13--016}
\preprint{CFTP/13--007}

%%%%%%%%%%%%%%%%%%%%%%%%%%%%%%%%%%%%%%%%%%%%%%%%%%%%%%%%%%%%%%%%%%%%%%%
\title{Constraints on dark matter annihilation from CMB observations
  before Planck}

\author{Laura Lopez-Honorez$^1$, Olga Mena$^2$, Sergio
  Palomares-Ruiz$^{2,3}$ and Aaron C. Vincent$^2$} 
\affiliation{$^1$
Theoretische Natuurkunde, Vrije Universiteit Brussel and The
International Solvay Institutes Pleinlaan 2, B-1050 Brussels, Belgium}  
\affiliation{$^2$Instituto de F\'{\i}sica Corpuscular (IFIC),
  CSIC-Universitat de Val\`encia, \\  
Apartado de Correos 22085, E-46071 Valencia, Spain}
\affiliation{$^3$Centro de F\'{\i}sica Te\'{o}rica de Part\'{\i}culas
  (CFTP), Instituto Superior T\'{e}cnico, \\ 
       Universidade T\'{e}cnica de Lisboa, Av. Rovisco Pais 1,
       1049-001 Lisboa, Portugal}

\date{\today}
\begin{abstract}
We compute the bounds on the dark matter (DM) annihilation cross
section using the most recent Cosmic Microwave Background measurements 
from WMAP9, SPT'11 and ACT'10.  We consider DM with mass in the 
MeV--TeV range annihilating $100\%$ into either an $e^+e^-$ or a
$\mu^+\mu^-$ pair.  We consider a realistic energy deposition model,
which includes the dependence on the redshift, DM mass and
annihilation channel.  We exclude the canonical thermal relic
abundance cross section ($\sv = 3 \times 10^{-26} {\rm cm}^3 {\rm
  s}^{-1}$) for DM masses below 30~GeV and 15~GeV for the $e^+e^-$
and $\mu^+\mu^-$ channels, respectively.  \textit{A priori}, DM
annihilating in halos could also modify the reionization history of
the Universe at late times.  We implement a realistic halo model taken
from results of state--of--the--art N--body simulations and consider a
mixed reionization mechanism, consisting on reionization from DM as
well as from first stars.  We find that the constraints on DM
annihilation remain unchanged, even when large uncertainties on the
halo model parameters are considered. 
\end{abstract}
\pacs{}
\maketitle

\section{Introduction}

The cosmic microwave background (CMB) radiation not only encodes
information about the early universe until the epoch of recombination,
but also about the integrated history of the intergalactic medium (IGM)
between the surface of last scattering and the present. At early
times, energy released into the IGM by dark matter (DM) annihilation
affects the recombination process, increasing the ionization fraction
and broadening the last scattering surface.  At late times,
annihilating DM in halos could also modify the reionization mechanism of
the Universe.  Therefore, DM annihilations generally alter the thermal
history of the Universe, leading to observable changes in CMB
temperature and polarization spectra~\cite{Chen:2003gz, Hansen:2003yj,
  Pierpaoli:2003rz, Padmanabhan:2005es, Mapelli:2006ej, Zhang:2006fr,
  Ripamonti:2006gq, Chuzhoy:2007fg, Finkbeiner:2008gw, Natarajan:2008pk,
  Natarajan:2009bm, Belikov:2009qxs, Galli:2009zc, Slatyer:2009yq,
  Hutsi:2009ex, Cirelli:2009bb, Kanzaki:2009hf, Chluba:2009uv,
  Valdes:2009cq, Natarajan:2010dc, Galli:2011rz, Hutsi:2011vx,
  Evoli:2012zz, Giesen:2012rp, Evoli:2012qh, Slatyer:2012yq,
  Frey:2013wh, Cline:2013fm, Weniger:2013hja} and to marginally
observable effects in the CMB bispectrum~\cite{Dvorkin:2013cga}.

New CMB data have recently become available.  The Wilkinson Microwave
Anisotropy Probe (WMAP) collaboration has made publicly available
their final nine--year data release~\cite{Hinshaw:2012fq}.  The South
Pole Telescope (SPT) collaboration has released their measurements of
2540~deg$^2$ of sky, providing the CMB temperature anisotropy power
over the multipole range
$650<\ell<3000$~\cite{Hou:2012xq,Story:2012wx}, in the region from the
third to the ninth acoustic peak.  Finally, the Atacama Cosmology
Telescope collaboration (ACT) has also released new measurements of
the CMB damping tail~\cite{Sievers:2013wk}.
 
In this work, we compute the constraints from these recent CMB
measurements on annihilating DM particles with masses ranging from
1~MeV to 1~TeV, in combination with Baryon Acoustic Oscillation (BAO)
data and Hubble Space Telescope (HST) measurements of the Hubble
constant.  Since ACT and SPT high multipole CMB measurements seem to
disagree on the value of several cosmological parameters such as the
effective number of relativistic species, the lensing potential, the
neutrino masses and the properties of the dark radiation
background~\cite{DiValentino:2013mt, Calabrese:2013jyk,
  Archidiacono:2013lva}, we analyze these two data sets separately.
  
We focus on purely leptonic annihilation channels, as $e^+e^-$ and
$\mu^+ \mu^-$.  Whereas in the case of the former channel very
efficient energy injection into the IGM would occur, in the case of
the latter an important part of the energy from DM annihilations would
be ``lost'' in the form of neutrinos.  Let us note that constraints on
DM annihilating into nearly all other\footnote{Constraints from 
  $\chi \chi \rightarrow \tau^+ \tau^-$  channel are only slightly
  stronger than the constraints on  $\chi \chi \rightarrow \mu^+
  \mu^-$   (see, e.g., Ref.~\cite{Cline:2013fm}).} two--particle SM
final states should fall between the bounds obtained for $\mu^+ \mu^-$
and $e^+e^-$.  These channels are also partly motivated by recent
explorations of astroparticle ``anomalies'', which point to a
significant fraction of positrons above the expected background.  The
annihilation of leptophilic DM has been a candidate explanation for
these anomalies, which include the positron fraction excess observed
by PAMELA and Fermi above 10~GeV~\cite{ArkaniHamed:2008qn,
  Pospelov:2008jd, Cholis:2008qq, Cholis:2008wq, Meade:2009iu,
  Chen:2009ab, Kane:2009if, Cirelli:2008pk, Kuhlen:2008aw,
  Feng:2010zp, Cline:2010ag, Vincent:2010kv}, the strong 511~keV
signal from the galactic center seen most recently by the INTEGRAL/SPI
experiment~\cite{Boehm:2003bt, Hooper:2003sh, Boehm:2004gt,
  Beacom:2004pe, Ascasibar:2005rw, Gunion:2005rw, Beacom:2005qv,
  Boehm:2006df, Sizun:2006uh, Finkbeiner:2007kk, Pospelov:2007xh,
  Prantzos:2010wi, Vincent:2012an}, and the large isotropic radio
signal below 10~GHz observed by ARCADE~\cite{Fornengo:2011cn,
  Fornengo:2011xk, Fornengo:2011iq, Hooper:2012jc, Yang:2012qi,
  Cline:2012hb}.

We therefore explore here DM annihilations into either pure $e^+e^-$
or pure $\mu^+ \mu^-$ final states, using a realistic redshift-- and DM
mass--dependent model of energy injection into the IGM.  We address
the impact of DM annihilations through the epochs of recombination and
reionization until today, considering a realistic halo formation
model, and compute the constraints on the DM mass and annihilation
cross section from the latest available cosmological data.

We begin by very briefly summarizing the theory of DM annihilation and
its impact on CMB photons, before presenting our methods and results.
Thus, in Section~\ref{sec:deposition}, we explicitly show the
parametrization of the energy injection and deposition from DM
annihilation, including the enhancement due to halo formation.
Section~\ref{sec:cosmo} summarizes the cosmological signatures of
extra injected energy into the IGM.  In Section~\ref{sec:analysis} we
provide the details of our analysis, and finally give the results of
our work in Section~\ref{sec:results}.  Details of the DM annihilation
spectra are presented in Appendix~\ref{sec:spectrum} and the
parametrization of the contribution from halos is given in
Appendix~\ref{sec:halos}.

\section{Energy deposition}
\label{sec:deposition}

In the following, we carefully distinguish between \textit{injected}
energy, defined as the energy liberated by DM annihilations, and
\textit{deposited} energy, which corresponds to the energy that 
actually goes into heating and ionization of the IGM.  The
annihilation of DM particles in the uniform density field of DM gives
rise to an injection power into the IGM per unit volume at redshift
$z$ given by
\begin{equation}
\left( \frac{\ud E}{\ud V \ud t}\right)_{\rm injected}=(1+z)^6
(\Omega_{\rm DM, 0}\rho_{c,0})^2\zeta \frac{\sv}{ m_{\chi}} ~, 
\label{eq:injectedEnergy}
\end{equation}
where $\Omega_{\rm DM, 0}$ is the present DM abundance, $\rho_{c,0}$ is
the critical density today and $\langle \sigma v \rangle$ is the
thermally--averaged DM annihilation cross section.  If DM particles
and antiparticles are identical $\zeta$, the counting factor, is $\zeta
=1$ (which we will use in this work), and $\zeta =1/2$ otherwise.  If
all the DM energy went into reheating the IGM,
Eq.~(\ref{eq:injectedEnergy}) would also correspond to the 
\textit{deposited} energy.  However, there are two factors which affect 
the efficiency of this process: 1) creation of ``invisible'' products,
such as neutrinos as daughter particles; and 2) the free--streaming of
electrons and photons from their creation until their energy is
completely deposited into the IGM via photoionization, Coulomb
scattering, Compton processes, bremsstrahlung and recombination.  If
this deposition is not complete, whatever remains would  contribute to
the present isotropic radiation background.  Process 1) depends on the
annihilation model, whereas process 2) depends on density and
redshift, and is completely specified by the history of the
IGM~\cite{Ripamonti:2006gq, Slatyer:2009yq, Valdes:2009cq,
  Slatyer:2012yq, Evoli:2012zz}.  It is therefore convenient to
parametrize the deposited energy as~\cite{Slatyer:2009yq,
  Slatyer:2012yq}
\begin{equation}
\left(\frac{\ud E}{\ud t \ud V} \right)_{\mathrm{deposited}} = 
f(z,\mx) \left(\frac{\ud E}{\ud t \ud V} \right)_{\mathrm{injected}},
\label{fofzdef}
\end{equation}
so that all the $z$--dependence is in the factor $f(z,\mx)(1+z)^6$,
where $f(z,\mx)$ represents the fraction of the injected energy by
annihilations of DM particles with mass $\mx$ which is deposited into
the IGM at redshift $z$.  This efficiency factor\footnote{Our
  $f(z,\mx)$ is identical to the $f(z)$ defined by
  Ref.~\cite{Slatyer:2012yq}.  We have made the $\mx$ dependence
  explicit for clarity.} is defined as~\cite{Slatyer:2012yq}
\begin{equation}
f(z,\mx) = \frac{H(z)}{(1+z)^3\sum_i \int \ud E \, E \, \frac{\ud N_i
    (E, \mx)}{\ud E}} \sum_i \int \ud z' \frac{(1+z')^2}{H(z')}\int
\ud E \, 
T_i(z',z,E) \, E \, \frac{\ud N (E, \mx)}{\ud E} ~.  
\label{fofz}
\end{equation}
The function $T_i(z',z,E)$ encodes the efficiency with which a photon
($i = \gamma$) or an electron/positron ($i = e^\pm$) injected at
redshift $z'$ and with energy $E$ is deposited into the IGM at
redshift $z$.  The factors of $(1+z')$ in the integral are due to:  
the DM annihilation rate which goes as $ n_{\rm DM}^2 \propto (1+z')^6$,
the volume element $\ud V$ which is proportional to $(1+z')^{-3}$, and
the conversion from injection per unit time to injection per unit
redshift, $\ud t = - \ud \ln(1+z')/H(z')$.  The spectrum of injected
particles, $dN/dE$, depends on the DM mass $\mx$ and on the
annihilation channel.  The denominator of Eq.~(\ref{fofz}) ensures that 
$f(z,\mx)$ is properly normalized to Eq.~(\ref{eq:injectedEnergy}).
The transfer function $T(z',z,E)$ was computed by several
authors~\cite{Slatyer:2009yq, Valdes:2009cq, Slatyer:2012yq,
  Evoli:2012zz} and has been tabulated and made 
public\footnote{\url{http://nebel.rc.fas.harvard.edu/epsilon/}}
by Slatyer~\cite{Slatyer:2012yq}.  

It is interesting to note that the energy deposition efficiency can be
larger than one for certain values of the DM mass.  This is because
particles injected into the ``transparency window'', $10^6
\,\mathrm{eV} \lesssim E \lesssim 10^{12} \,\mathrm{eV}$, may
propagate freely \cite{Chen:2003gz, Padmanabhan:2005es,
  Slatyer:2009yq} until they are finally redshifted to energies low
enough to efficiently Compton--scatter.  This appears as an ``extra''
source of energy at lower $z$ than one would not expect in the
on--the--spot approximation, where energy deposition is taken to be
instantaneous\footnote{The on--the--spot approximation, used in several
  previous analyses~\cite{Chen:2003gz, Hansen:2003yj,
    Pierpaoli:2003rz, Padmanabhan:2005es, Mapelli:2006ej, Zhang:2006fr,
    Finkbeiner:2008gw, Galli:2009zc}, is recovered with $T(z',z,E) =
  f \, \delta\left(\ln(1+z')-\ln(1+z)\right)$, where $f$ is a constant
  efficiency factor, formally different from our $f_{\rm eff}$ defined
  below.}.  For practical reasons, we define below an effective energy
deposition efficiency $\feff$ which depends on the DM mass.

\subsection{Inclusion of DM halos}

Since the injection power from DM annihilation goes as $n_{\rm DM}^2$,
the enhancement due to the DM number density at late times by the
formation and collapse of bound structures can provide a significant
enhancement to the rate of injected energy.  In order to compute the
halo contribution, the halo mass function $\ud n_{\rm halo} / \ud M$
is required to specify the number of halos as a function of the halo
mass $M$ and redshift $z$.  Some works~\cite{Natarajan:2008pk,
  Natarajan:2009bm, Cirelli:2009bb, Kanzaki:2009hf, Giesen:2012rp}
have used the analytic Press--Schechter spherical collapse
model~\cite{Press:1973iz}, whereas others~\cite{Belikov:2009qxs,
  Hutsi:2009ex, Natarajan:2010dc, Hutsi:2011vx, Natarajan:2012ry} have
considered the Sheth--Tormen ellipsoidal model~\cite{Sheth:1999mn}.
In this work we follow a fully numerical approach by using the results
of N--body simulations to obtain the halo mass
function~\cite{Watson:2012mt} and take the relation of the concentration
parameter to the halo mass from Ref.~\cite{Prada:2011jf}, which we
implement to obtain the single halo contribution assuming a Navarro,
Frenk and White (NFW) DM density profile~\cite{Navarro:1995iw}.  We
provide the details of the used inputs, as well as some useful
analytical parametrizations, in Appendix~\ref{sec:halos}.  The energy 
injected into the IGM by DM annihilation in halos at redshift $z$ is
given by 
\begin{equation}
\left(\frac{\ud E}{\ud V \ud t}\right)_{\rm halo, injected}  =  \zeta \,
\frac{\langle \sigma v \rangle}{m_\chi} \, \times \,
\int_{M_\mathrm{min}}^\infty \ud M \frac{\ud n_{\rm halo} (M,z)}{\ud M} 
 \, \int_0^{r_{\Delta}} \ud r \, 4 \pi r^2 \, \rho_{\rm halo}^2(r) ~,
\label{eq:haloinjectedE}
\end{equation}
where the second integral represents the single halo contribution and
the first integral parametrizes the weighted sum over all halos.  The
density profile of a DM spherical halo of mass $M$, radius
$r_{\Delta}$ and concentration parameter $c_\Delta(M,z)$ is $\rho_{\rm 
  halo}(r)$.  In this work we use $\Delta = 200$.  These quantities
are defined so that $M = 4\pi/3 \, r_{\Delta}^3 \, \Delta \, \rho_{\rm
  c} (z)$.  As for the lower limit of integration in the first integral,
it has been shown that the two main damping processes that generate a
exponential cutoff in the power spectrum are the free--streaming of DM
particles from high to low density regions~\cite{Green:2005fa} and the
effect of acoustic oscillations in the cosmic bath~\cite{Loeb:2005pm,
  Bertschinger:2006nq}.  These processes depend on the particle
physics and cosmological model and hence the minimum halo mass $M_{\rm
  min}$ is a very model dependent quantity that can vary from $M_{\rm
  min} = 10^{-4} M_\odot$ to $M_{\rm min} = 10^{-11}
M_\odot$~\cite{Bringmann:2009vf, Cornell:2012tb, Gondolo:2012vh}.
Observational probes of these microhalos have been studied in the
literature~\cite{Pieri:2005pg, Koushiappas:2006qq, Bringmann:2009vf,
  Cornell:2012tb, Gondolo:2012vh}.  Here we use $M_{\rm min} = 10^{-6}
M_\odot$ and although there is significant uncertainty on the order of
magnitude of $M_{\rm min}$, the total deposited energy only depends
weakly on it.

Obtaining the energy deposited by the halo component of the DM
requires performing the same integral as in Eq.~(\ref{fofz}), over the  
transfer matrix $T(z',z,E)$.  For simplicity, we keep the same
parametrization as before, separating out the $z$--dependence.  The
total injection power is therefore given by 
\begin{equation}
\left(\frac{\ud E}{\ud t \ud V} \right)_{\mathrm{deposited}} = \left[
  f(z,\mx) + g(z,\mx) \right] (1+z)^6 (\Omega_{\rm DM, 0} \, \rho_{\rm
  c, 0})^2 \, \zeta \, \frac{\langle  \sigma v \rangle}{ m_{\chi}} ~, 
\end{equation}
where $g(z,\mx)$, the part of the efficiency factor due to DM halos, is
\begin{equation}
g(z,\mx) = \frac{H(z)}{(1+z)^3\sum_i \int E \frac{\ud N}{\ud E}\ud E}
\sum_i \int \ud z' \, \frac{(1+z')^2}{H(z')} \, G(z') \, \int
T_i(z',z,E) \, E \, \frac{\ud N}{\ud E} \ud E ~,   
\label{eq:gofz}
\end{equation}
and the dimensionless function $G(z)$ is defined as
\begin{equation}
G(z) \equiv \frac{1}{\left(\Omega_{\rm DM, 0} \, \rho_{c,0}\right)^2}
\, \frac{1}{(1+z)^6} \, \int \ud M \, \frac{\ud n(M,z)}{\ud M}  \,
\int_0^{r_{\Delta}} \ud r \, 4 \pi r^2 \, \rho_{\rm halo}^2(r) ~. 
\label{eq:G}
\end{equation}
We also provide a parametrization of the function $g(z,\mx)$ for the
case of DM annihilation into $e^+ e^-$ pairs in
Appendix~\ref{sec:halos}.

\section{Cosmological signatures of DM annihilations}
\label{sec:cosmo}

The electromagnetic energy released by DM annihilations would be
deposited into the IGM, inducing heating as well as ionizations and
excitations of hydrogen and helium atoms.  Excited atoms have a
certain probability of being subsequently
ionized~\cite{Peebles:1968ja}.  Nevertheless, in this work we do not  
consider this extra source of ionization, which is expected to have a
relatively weak effect, of the order of $\sim 10\%$~\cite{Galli:2011rz,
  Hutsi:2011vx}.  In order to compute the changes in the ionization
history we use the CosmoRec package~\cite{Chluba:2010ca,
  Chluba:2009uv, AliHaimoud:2010ab, Chluba:2010fy, Grin:2009ik,
  Switzer:2007sq, RubinoMartin:2009ry} that includes a subroutine that
modifies the evolution equations for the IGM temperature and for the
net ionization rate from the ground states of neutral hydrogen and
helium as
\begin{eqnarray}
\left.\frac{\ud T_{\rm m}}{\ud z}\right|_{\rm DM} & = & - \,
\frac{1}{(1+z) \, H(z)} \, \frac{2}{3k_{\rm B}} \, \frac{\chi_{\rm
    h}(z)}{N_{\rm H}(z) \, [1 + f_{\rm He} + x_{\rm e} (z)]} \,
\left(\frac{\ud E}{\ud t \ud V}\right)_{\mathrm{deposited}}
~; \label{eq:recomb1} \\ 
\left.\frac{\ud x_{\rm p}}{\ud z} \right|_{\rm DM}
& = & \frac{1}{(1+z) \, H(z)} \, \frac{1}{N_{\rm H}(z) \, [1 + f_{\rm
      He}]} \, \frac{\chi_{\rm p}(z)}{E^{\ion{H}{i}}_{\rm ion}}\,
\left(\frac{\ud E}{\ud t \ud V} 
\right)_{\mathrm{deposited}} ~; \label{eq:recomb2}\\  
\left.\frac{\ud x_{\ion{He}{ii}}}{\ud z} \right|_{\rm DM} & = &
\frac{1}{(1+z) \, H(z)} \, \frac{f_{\rm He}}{N_{\rm H}(z) \, [1
    +f_{\rm He}]} \, \frac{\chi_{\rm He}(z)}{E^{\ion{He}{i}}_{\rm
    ion}} \, \left(\frac{\ud E}{\ud t \ud V}
\right)_{\mathrm{deposited}} ~, \label{eq:recomb3} 
\end{eqnarray}
where $T_{\rm m}$ is the IGM temperature, $x_{\rm p} \equiv N_{\rm
  p}/N_{\rm H}$ ($x_{\ion{He}{ii}} \equiv N_{\ion{He}{ii}}/N_{\rm H}$)
is the fraction of ionized atoms from the ground states of neutral
hydrogen (helium) relative to the total number of hydrogen nuclei  
$N_{\rm H}(z)$, $f_{\rm He} \equiv N_{\rm He}/N_{\rm H} = 0.0795$ is the
fraction of helium nuclei, $x_{\rm e} \equiv N_{\rm e}/N_{\rm H}
\simeq x_{\rm p} + x_{\ion{He}{ii}}$ is the free electron fraction,
$k_{\rm B}$ is the Boltzmann constant and $E^{\ion{H}{i}}_{\rm ion} =
13.6$~eV and $E^{\ion{He}{i}}_{\rm ion} = 24.6$~eV are the ionization
potentials for hydrogen and helium, respectively.  The specific
contributions to the fraction of injected energy converted to
ionization from hydrogen and helium are given by $\chi_{\rm H}(z)$ and
$\chi_{\rm He}(z)$ and that converted into heat is $\chi_{\rm h}(z)$.
It has been shown that for a neutral plasma the amount of energy that
goes into heat, ionization and excitation is roughly the same, whereas
if it is fully ionized all the energy gets converted into
heat~\cite{1985ApJ...298..268S}.  This suggested a simple
approximation~\cite{Chen:2003gz}, further refined by including the
ionization of helium~\cite{Padmanabhan:2005es},
\begin{eqnarray}
\chi_{\rm h}(z) & \approx & \frac{1 + 2 \, x_{\rm p}(z) + f_{\rm He}
  (1 + 2 \, Z_{\ion{He}{ii}}(z))}{3 \, (1 + f_{\rm He})} = \frac{1}{3}
\, \left( 1 + 2 \, \frac{x_{\rm e} (z)}{1 + f_{\rm He}}\right)~;\\
\chi_{\rm p} (z) & \approx & \frac{1 - x_{\rm p}(z)}{3} ~,\\
\chi_{\rm He}(z) & \approx & \frac{1 - Z_{\ion{He}{ii}} (z)}{3}~.
\end{eqnarray}
where $Z_{\ion{He}{ii}} = N_{\ion{He}{ii}}/N_{\rm He}$ is the fraction of
singly ionized helium atoms relative to the total number of helium 
nuclei.

\begin{figure}[t]
\begin{tabular}{cc}
\includegraphics[width=.7\textwidth]{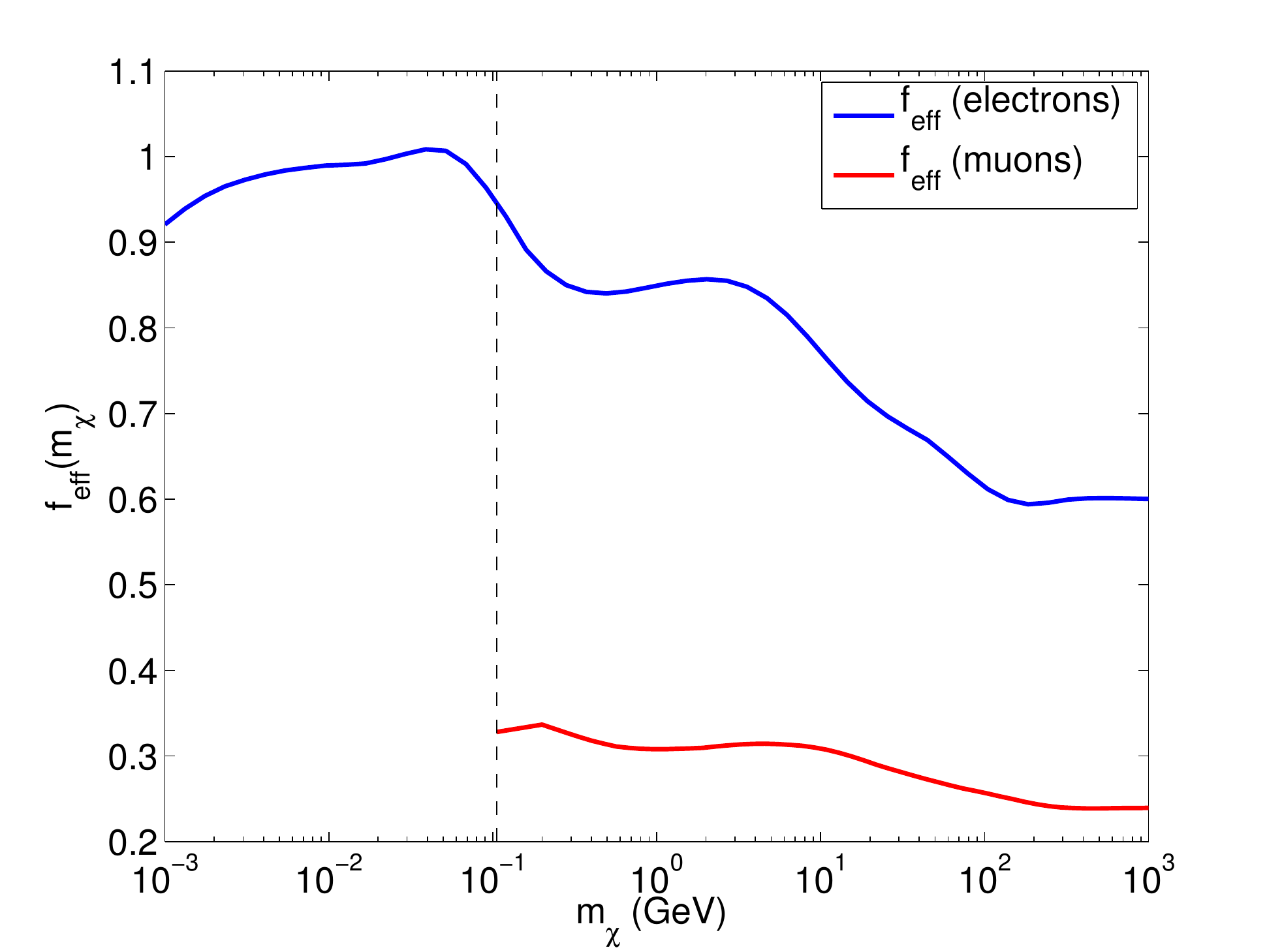}
  \end{tabular}
  \caption{\sl The effective energy deposition fractions for the
    smooth DM background component, as a function of the DM mass, in
    GeV.  Blue (upper) and red (lower) curves denote the $e^+e^-$ and
    $\mu^+\mu^-$ annihilation channels explored in this study.  Red
    (lower) curve stop at the muon mass, the threshold DM mass for muon
    production via annihilation.}
  \label{fig:feff_geff}
\end{figure}

\begin{figure}
\includegraphics[width=.7\textwidth]{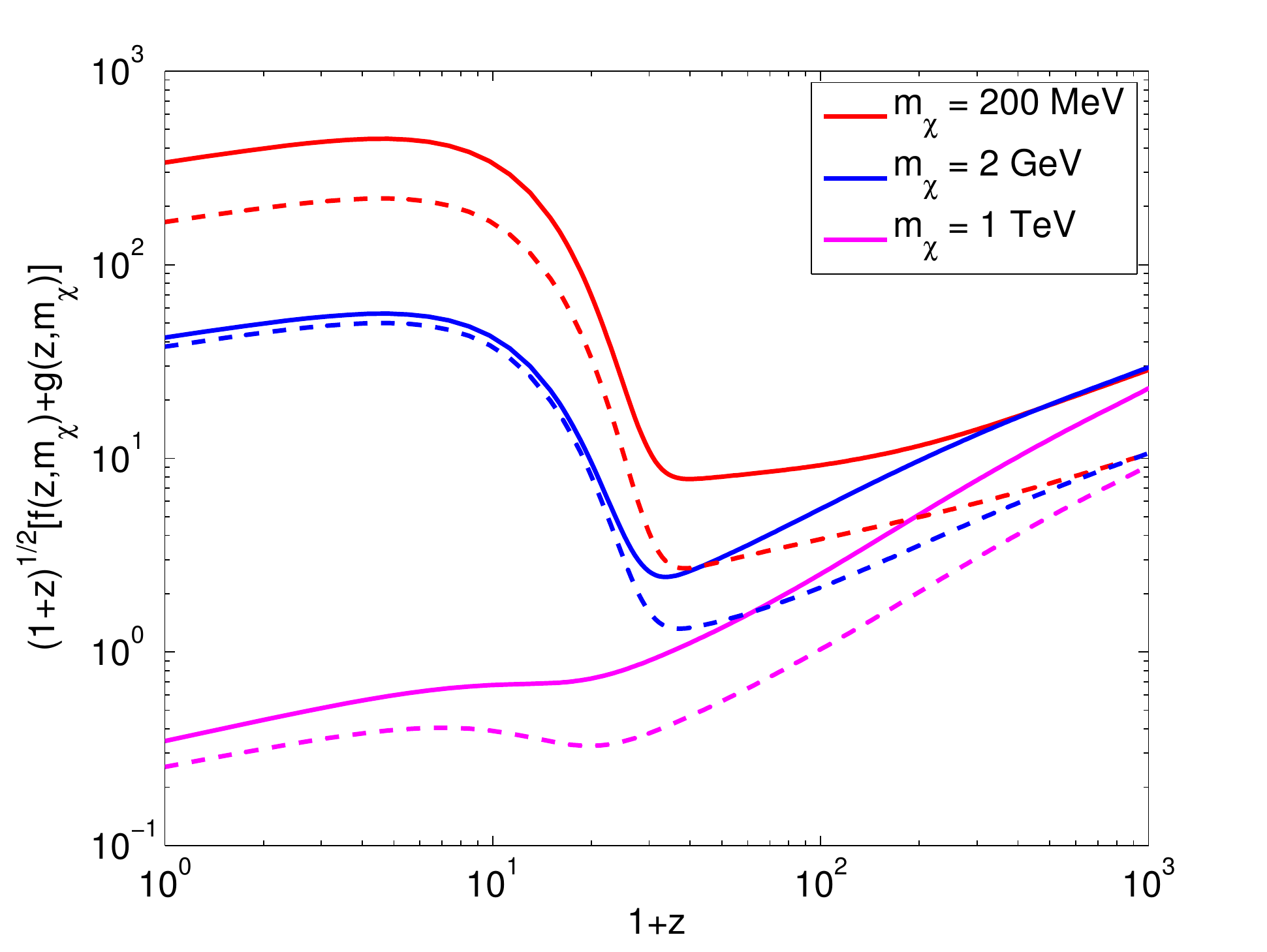}
\caption{\sl The redshift--dependence of the extra contribution to
  heating and ionization due to DM annihilation, $\sqrt{1+z} \, [f(z,
    \mx) + g(z, \mx)]$, for three values of the DM mass (from top to
  bottom, $\mx = 200$~MeV, 2~GeV and 1~TeV).  Solid lines are for
  $\chi \chi \rightarrow e^+ e^-$ and dashed lines for the $\chi \chi
  \rightarrow \mu^+\mu^-$ channel.  The steep rise at low redshifts is
  due to halo formation.}
\label{fig:zdep}
\end{figure}

Now, with these modifications to the evolution equations due to DM
annihilations, we can define an effective energy deposition efficiency
$\feff(\mx)$ (for the contribution from the smooth DM component) in
order to bypass the computationally expensive interpolation at each
redshift of $f(z,\mx)$ in our Monte Carlo analyses,
\begin{equation}
\feff (\mx) = \frac{ \displaystyle \int_{\zmax}^{\zmin} f(z,\mx)
  \sqrt{1+z} ~ \ud z}{\displaystyle \int_{\zmax}^{\zmin} \sqrt{1+z} \, 
  \ud z} ~,
\label{eq:feff}
\end{equation}
and it parametrizes the effective value of $f(z,\mx)$ through the
history of the Universe.  The factor $\sqrt{1+z}$ is motivated by the
redshift dependence of Eqs.~(\ref{eq:recomb1})--(\ref{eq:recomb3}),
which contain the terms $n_{\rm DM}^2 \propto (1+z)^6$, $N_{\rm{H}}^{-1}
\propto (1+z)^{-3}$ and $((1+z) H(z))^{-1} \sim (1+z)^{-5/2}$ (in the
matter--dominated regime).  The lower limit for integration is set at
$\zmin = 50$, before reionization starts and the upper limit is chosen
to be $\zmax=1100$, when the effect of DM annihilations starts to
delay recombination\footnote{We have checked that using $\zmin<50$
  or $\zmax>1100$ only changes our results by a few percent.}.  We have
verified that the results obtained with the effective deposition
efficiency $\feff(\mx)$ differ by less than a few percent from those
obtained with the full redshift--dependent $f(z,\mx)$.  

Fig.~\ref{fig:feff_geff} shows $\feff(\mx)$ for DM annihilations into
$e^+e^-$ and $\mu^+\mu^-$, as a function of the DM mass.  We note that
the effective $\feff(\mx)$ deposition efficiency is much lower for the
$\chi \chi \rightarrow \mu^+\mu^-$ channel due the large fraction of
final--state neutrinos, which do not heat the IGM nor contribute to
ionization.  As noted in Section~\ref{sec:deposition}, $f(z,\mx)$ and
hence $\feff(\mx)$, is larger than 1 for a certain mass range ($\mx
\sim 100$~MeV) due to a ``pile--up'' effect: photons and
electrons/positrons emitted at early times into their transparency 
window are eventually redshifted enough to efficiently deposit 
their energy at later times.  Conversely, energy injected from DM
particles with masses in the GeV--TeV region will remain in the
``transparency window'' until the present time lowering the
$\feff(\mx)$ injection efficiency.
  
In addition to $\feff(\mx)$, we also use an approximation for $g(z,
\mx)$ so that the $z$-- and $\mx$--dependences are factored out,
\begin{equation}
\geff (z, \mx) = \gamma(\mx) \, \Gamma(z) 
\label{eq:geff}
\end{equation}
The fitted functions $\gamma(\mx)$ and $\Gamma(z)$ are qualitatively
similar to the shape of $\feff(\mx)$, and are given in
Appendix~\ref{sec:halos}.

The total deposited energy that goes into the ionization equations is
therefore
\begin{equation}
\left(\frac{\ud E}{\ud t \ud V} \right)_{\mathrm{deposited}} = \left(
\feff(\mx) + \geff(z, \mx) \right) \, \left(\frac{\ud E}{\ud t \ud V}
\right)_{\mathrm{injected}} ~. 
\end{equation}
Once this is taken into account, the redshift--dependence of the extra
ionization and heating terms becomes: $\sim \sqrt{1+z} \left[f(z,\mx) +
  g(z,\mx)\right]$. This function is shown for three values of $\mx$
in Fig.~\ref{fig:zdep}.

\begin{figure}[t]
\begin{center}
\includegraphics[width=.6\textwidth]{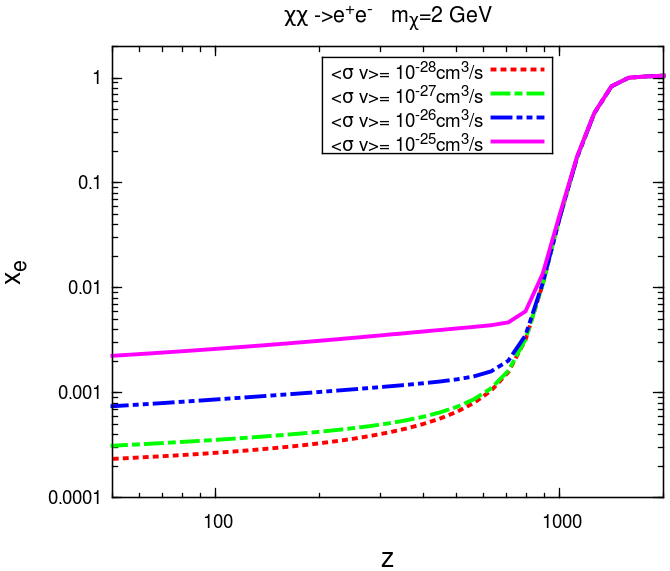}
\end{center}
  \caption{\sl Free electron fraction, $x_{\rm e}(z)$, as a function
    of the redshift $z$, for $\mx = 2$~GeV and DM annihilations into
    an $e^+e^-$ pair for different values of the annihilation cross
    section.}
  \label{fig:xe}
\end{figure}

The main effect of the extra injection of energy from DM annihilations
is the delay in recombination at $z\sim 1100$ and an enhancement in
the low--redshift tail of the ionization fraction.  This modifies the
optical depth for CMB photons as they travel from the last scattering
surface to us, so that the visibility function, the probability that a
photon last scattered at a given redshift, extends to lower redshifts.
This changes the position of the acoustic peaks in the temperature
power spectrum and broadens the surface of last scattering,
suppressing perturbations on scales smaller than the width of this
surface and thus attenuates the power spectrum.  In addition, the
amplitude of the polarization fluctuations is increased and the
positions of the TE and EE peaks are shifted, whereas at smaller
scales these power spectra are attenuated, similarly to what happens
for the temperature power spectrum (see, e.g., Refs~\cite{Chen:2003gz,
  Padmanabhan:2005es}). 

The impact of annihilating DM on the CMB spectrum is thus an integrated
effect which lasts from the early recombination era (high redshift, $z
\sim 1100$) until the reionization epoch at late times (low redshift,
$z \lesssim 10$).  In the early recombination period the presence of
the extra heating and ionization terms from DM annihilation processes
would change the free electron fraction below $z \sim 1100$.  This is
depicted in Fig.~\ref{fig:xe}, where we have chosen a DM mass of $\mx
= 2$~GeV for illustration.  When including the effects from the
annihilation of DM, the optical depth due to Thomson scattering increases,
broadening the last scattering surface and producing a damping of the
acoustic oscillations.  Although this effect is degenerate with the
slope and amplitude of the primordial perturbations, the polarization
spectrum helps disentangle the two effects.  A number of studies have
placed constraints on the DM masses and annihilating cross
sections/decaying rates by analyzing these signatures on the CMB
temperature and polarization spectra~\cite{Chen:2003gz, Hansen:2003yj,
  Pierpaoli:2003rz, Padmanabhan:2005es, Mapelli:2006ej, Zhang:2006fr,
  Ripamonti:2006gq, Chuzhoy:2007fg, Finkbeiner:2008gw, Natarajan:2008pk,
  Natarajan:2009bm, Belikov:2009qxs, Galli:2009zc, Slatyer:2009yq,
  Hutsi:2009ex, Cirelli:2009bb, Kanzaki:2009hf, Chluba:2009uv,
  Valdes:2009cq, Natarajan:2010dc, Galli:2011rz, Hutsi:2011vx,
  Evoli:2012zz, Giesen:2012rp, Evoli:2012qh, Slatyer:2012yq,
  Frey:2013wh, Cline:2013fm, Weniger:2013hja}.

At late times, annihilating DM in halos may also change the
reionization history, leaving an imprint in the large scale
polarization spectra.  Cosmological constraints arising from the 
reionization period on annihilating or decaying DM have been computed
in Refs.~\cite{Pierpaoli:2003rz, Mapelli:2006ej, Natarajan:2008pk,
  Natarajan:2009bm, Hutsi:2009ex, Cirelli:2009bb, Kanzaki:2009hf,
  Belikov:2009qxs, Natarajan:2010dc, Hutsi:2011vx, Giesen:2012rp}.
Although there have been attempts to explain the reionization of the
universe solely by the effect of DM annihilation~\cite{Chuzhoy:2007fg,
  Natarajan:2008pk, Natarajan:2009bm, Belikov:2009qxs, Hutsi:2009ex,
  Natarajan:2010dc, Giesen:2012rp}, we do not consider such a
possibility here.  We consider mixed reionization scenarios including
reionization from both DM annihilation in halos and first stars.  The
latter is accounted for by using the default reionization model
implemented in CAMB at a given redshift $z_{\rm reio}$.

\begin{figure}[t]
  \begin{tabular}{cc}
\includegraphics[width=0.5\textwidth]{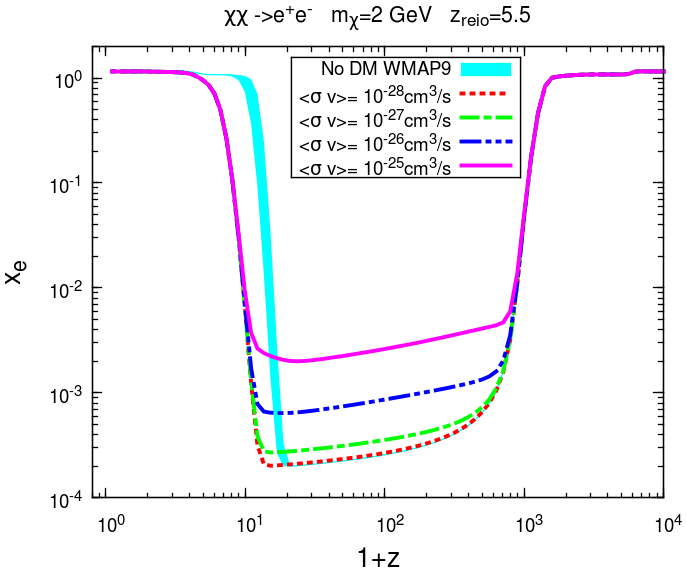}&
\includegraphics[width=0.5\textwidth]{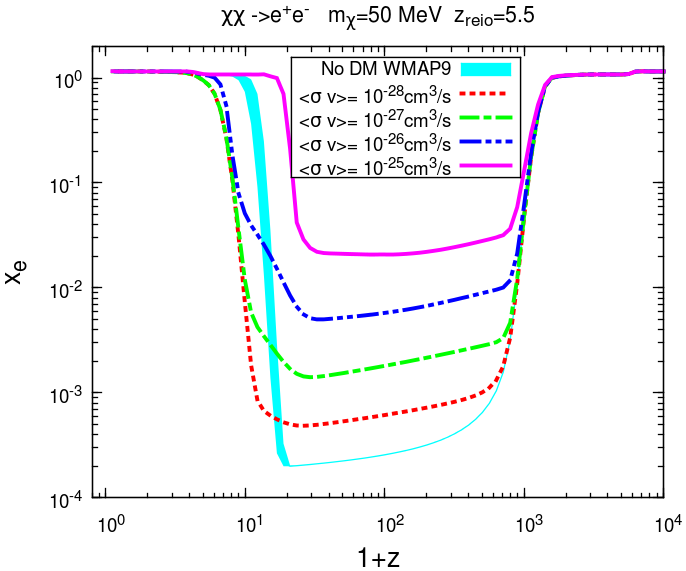}\\
  \end{tabular}
  \caption{\sl Free electron fraction, $x_{\rm e}(z)$, as a function
    of the redshift $z$, in the mixed model explored here, which
    considers reionization processes from both DM annihilation and
    first stars.  In the left and right panels we depict the cases of
    DM annihilation into $e^+ e^-$ with masses $\mx = 2$~GeV and $\mx
    = 50$~MeV, respectively.  In these scenarios, the reionization
    redshift in the CAMB simplified model for the contribution from
    first stars has been fixed to $z_{\rm reio} = 5.5$.  The $1\sigma$
    error band around the best--fit value from WMAP9 for the case of
    no contribution from annihilating DM is also plotted.}
  \label{fig:xe2}
\end{figure}

The free electron fraction as a function of the redshift for different
scenarios with and without DM contribution is depicted in
Fig.~\ref{fig:xe2}.  In both panels, the light blue band shows the
WMAP9 $1\sigma$ error around the best--fit value (in the absence of DM
contribution) for the optical depth to reionization $\tau=0.089\pm
0.014$.  In the left (right) panel of Fig.~\ref{fig:xe2}, we consider
mixed reionization scenarios with a contribution from the simplified
model for reionization from stars at $z_{\rm reio}=5.5$ plus a
contribution from DM particles of $\mx = 2$~GeV (50~MeV) annihilating
$100\%$ into an $e^+e^-$ pair.  We see that the impact of DM
contribution on reionization is rather suppressed. In order to get a 
non-negligible effect we have to consider small masses dark matter
masses, such as $\mx$ = 50~MeV represented here, and annihilation
cross sections as large as $\sv= 10^{-25}$ cm$^3$/s.  Given the halo
model considered in this work, we can not actually account for the
full reionization of the Universe at low redshift with MeV--TeV DM
taking into account the bounds on $\sv$ presented below in
sec.~\ref{sec:results}.

It has also been argued that annihilating DM in halos can increase the
reionization optical depth even if reionization by stars occurs at $z
\sim 6$, alleviating the tension between the value of the reionization
optical depth measured by CMB experiments and the fraction of neutral
hydrogen $x_{\rm H} = 1 - x_{\rm p}$ determined by observations of
Lyman--$\alpha$ absorption lines in quasar spectra (Gunn--Peterson 
effect~\cite{Gunn:1965hd}), which require that $x_{\rm H}\geq10^{-3}$,
and perhaps as high as 0.1 at $z \geq 6$~\cite{Fan:2006dp,
  Caruana:2012ww} and $x_{\rm H} \leq 10^{-4}$ at $z \leq 
5.5$~\cite{Fan:2006dp}.  This represents an abrupt change of $x_{\rm 
  H}$ (or equivalently $x_{\rm e}$) at $z \sim 6$, which cannot be
reproduced if $z_{\rm reio} \sim 10$, as indicated by CMB observations.
Therefore, and as pointed out in Refs.~\cite{Chuzhoy:2007fg,
  Natarajan:2008pk, Natarajan:2009bm, Belikov:2009qxs, Hutsi:2009ex, 
  Natarajan:2010dc, Giesen:2012rp}, the contribution from DM
annihilations in halos could, in principle, explain the measured
optical depth by CMB observations, while reionization from the first
stars at $z\sim 6$ could complete the reionization process and explain
the Gunn--Peterson bounds.  However, the value of the annihilation
cross section required for this reconciliation is badly excluded by
our CMB analyses.

\begin{figure}[t]
\begin{tabular}{cc}
\includegraphics[width=0.5\textwidth]{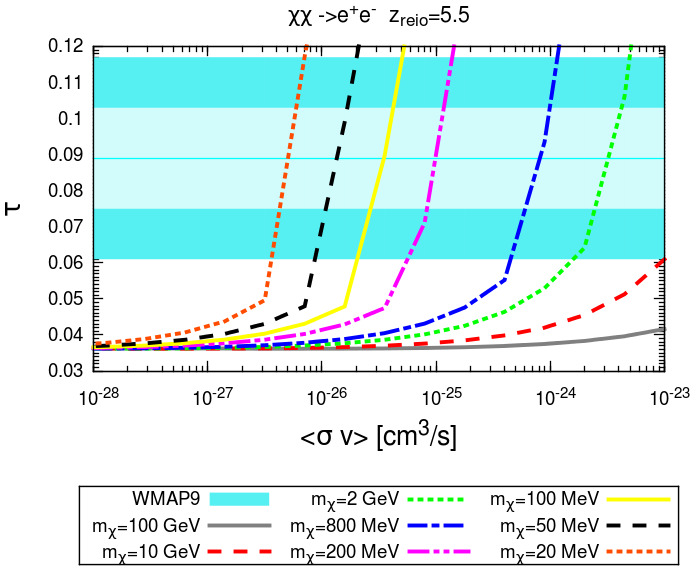}& 
\includegraphics[width=0.5\textwidth]{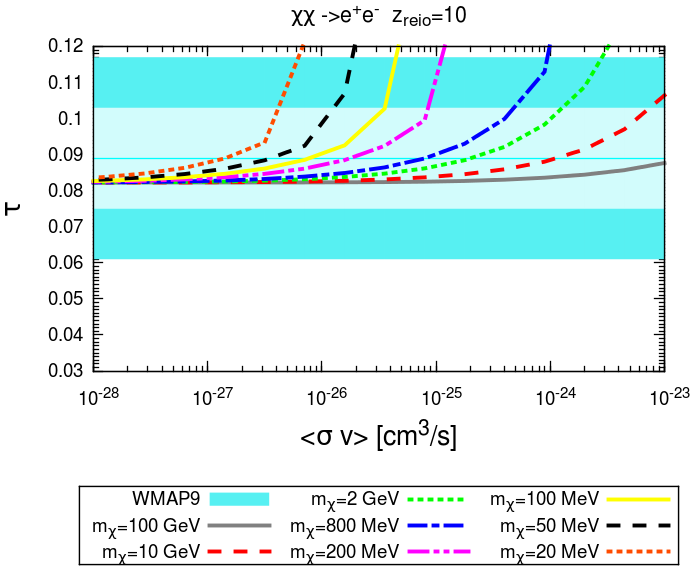}\\
  \end{tabular}
  \caption{\sl Reionization optical depth $\tau$, as a function
    of the DM annihilation cross section, in the mixed model explored
    here, which considers reionization processes from both DM
    annihilation and first stars.  We illustrate the behavior
    for different DM masses assuming $100\%$ annihilation into $e^+
    e^-$.  In these scenarios, the reionization redshift in the CAMB
    simplified model for the contribution from first stars has been
    fixed to $z_{\rm reio}=5.5$ (left panel) and 10 (right panel).
    The best--fit value and the $1\sigma$ and $2\sigma$ error bands 
    from WMAP9 for the case of no contribution from annihilating DM
    are also shown in light blue.}
  \label{fig:optdepth}
\end{figure}

The two panels of Fig.~\ref{fig:optdepth} depict the reionization optical depth\footnote{Notice that in order to compute the reionization optical depth $\tau$, we have integrated between redshift 100 (instead of redshift 40 which is  the default value in CAMB) and today.}
$\tau$ for different values of $\mx$ as a function of $\sv$ assuming two
different values of the redshift of reionization from stars,
$z_{\rm reio} = 5.5$ and 10.  We also show the $1\sigma$ and $2\sigma$
bands around the best--fit value for $\tau$ from WMAP9 data ($\tau =
0.089 \pm 0.014$)~\cite{Hinshaw:2012fq}.  From both panels, we can see
that the larger the DM mass the weaker the constraints on the
annihilation cross section, which is the expected behavior (see, e.g.,
Eq.~(\ref{eq:injectedEnergy})) due to a smaller number density of DM
particles for larger masses.  Notice that, for the case of the lower
reionization redshift $z_{\rm reio}=5.5$ (left panel of
Fig.~\ref{fig:optdepth}), as is well known, the optical depth without 
DM annihilation would be much lower than the best--fit value measured
by the WMAP team, lying three standard deviations away from it.  On the
other hand, DM masses in the 100's~MeV range with cross sections in the
$10^{-26} \, \textrm{cm}^3 \textrm{s}^{-1}$ could increase $\tau$ to
reach the observed value.  For the case of $z_{\rm reio}=10$
(right--hand panel of Fig.~\ref{fig:optdepth}), DM annihilations in  
halos are not needed in order to explain the measured value of $\tau$
and strong constraints can be placed on the DM annihilation cross
section.  In our numerical analyses, presented in the following
sections, we consider $z_{\rm reio}$ as a free parameter, to be
determined by the data.

\begin{figure}[t]
\bc
\vspace{0.5cm}
\includegraphics[width=0.5\textwidth]{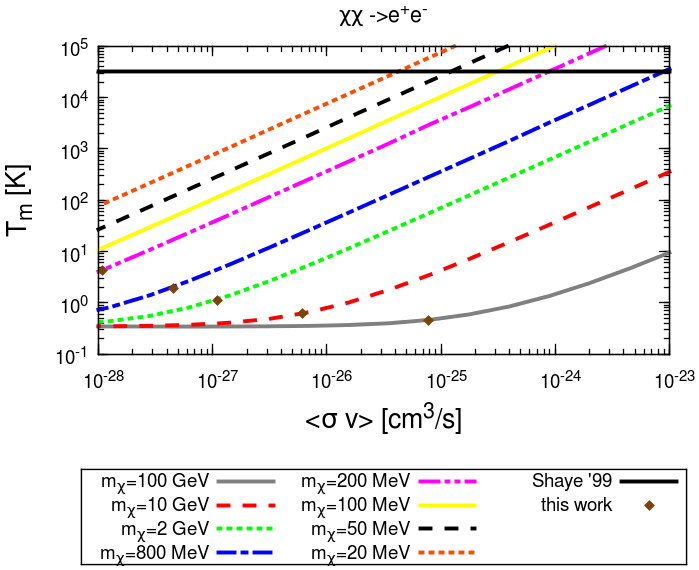} 
\ec
  \caption{\sl IGM temperature $T_{\rm m}$, in kelvin, as a function
    of the DM annihilation cross section for different DM masses,
    assuming that DM annihilation occurs $100\%$ into $e^+ e^-$ pairs.
    The reionization redshift is assumed to be $z_{\rm reio} = 10$.
    We indicate the value $T_{\rm m} = 32000$~K, upper limit from 
    Lyman--$\alpha$ observations for the IGM temperature in the
    $2<z<4.5$ redshift region~\cite{Schaye:1999vr}.  We also indicate
    the upper limit that we obtain for $\sv$ using
    WMAP9+SPT’11+HST+BAO data and the appropriate $f_{\rm eff}$ for  
    each DM mass with brown diamonds.}
  \label{fig:Tm}
\end{figure}

An additional constraint could arise from Lyman--$\alpha$
observations~\cite{Cirelli:2009bb, Giesen:2012rp}, which appear to
indicate that the IGM temperature is a few $10^{4}$\,K in the $2<z<4.5$
redshift region~\cite{Schaye:1999vr}.  Fig.~\ref{fig:Tm} depicts the
IGM temperature, $T_{\rm m}$ at redshift $z = 3$ for different values
of the annihilating DM mass $\mx$ versus $\sv$.  We indicate as well
the value $T_{\rm m}=32000$\,K, that has been considered in
Ref.~\cite{Giesen:2012rp} as a conservative upper bound on the IGM
temperature at low redshift.  The values of $\sv$ that saturate the
temperature bound are however several orders of magnitude above the
limits that we obtain from CMB data.  As a guide for the eye, we
reported the latter constraints with brown diamonds in
Fig.~\ref{fig:Tm}.

\section{Analysis}
\label{sec:analysis}
\subsection{Data and cosmological parameters}

In our analyses we explore cosmological scenarios that include DM
annihilations and are described by the following set of parameters: 
\begin{equation}
\label{eq:parameters}
  \{\omega_{\rm b},\omega_{\rm DM}, \Theta_{\rm s}, z_{\rm reio},
  n_{\rm s}, \log[10^{10}A_{\rm s}], \sv, \mx\} ~,
\end{equation}
where $\omega_{\rm b} \equiv \Omega_{\rm b, 0} h^{2}$ and $\omega_{\rm
  DM} \equiv \Omega_{\rm DM, 0}h^{2}$ are the physical baryon and cold
DM energy densities, $\Theta_{\rm s}$ is the ratio between the sound
horizon and the angular diameter distance at decoupling, $z_{\rm
  reio}$ is the reionization redshift, $n_{\rm s}$ is the scalar
spectral index, $A_{\rm s}$ is the amplitude of the primordial
spectrum, $\sv$ is the thermally--averaged DM annihilation cross
section and $\mx$ is the DM mass.  For our numerical analyses, we have
used the Boltzmann code CAMB~\cite{Lewis:1999bs}.  As explained above,
for the recombination calculation and DM contribution to reionization,
we have used the CosmoRec package~\cite{Chluba:2010ca, Chluba:2009uv,
  AliHaimoud:2010ab, Chluba:2010fy, Grin:2009ik, Switzer:2007sq,
  RubinoMartin:2009ry} since it contains subroutines that account for
the additional inputs relevant for the study of DM annihilations.  We
have then extracted cosmological parameters from current data using a
Monte Carlo Markov Chain (MCMC) analysis based on the publicly
available MCMC package \texttt{cosmomc}~\cite{Lewis:2002ah}.  Finally,
Table~\ref{tab:priors} specifies the priors considered on the
different cosmological parameters.

\begin{table}[t]
\begin{center}
\begin{tabular}{c|c}
\hline\hline
 Parameter & Prior\\
\hline
$\Omega_{\rm b, 0} h^2$ & $0.005 \to 0.1$\\
$\Omega_{\rm DM, 0} h^2$ & $0.01 \to 0.99$\\
$\Theta_{\rm s}$ & $0.5 \to 10$\\
$z_{\rm reio}$ & $6 \to 12$\\
$n_{\rm s}$ & $0.5 \to 1.5$\\
$\ln{(10^{10} A_{\rm s})}$ & $2.7 \to 4$\\
$\langle  \sigma v \rangle/(3\cdot 10^{-26} \textrm{cm}^3/\textrm{s})$
&  $10^{-5} \to 10^{2.5}$\\ 
\hline\hline
\end{tabular}
\caption{\sl Uniform priors for the cosmological parameters considered
  here and the DM annihilation cross section.} 
\label{tab:priors}
\end{center}
\end{table}

Our baseline data set is the WMAP9 data~\cite{Hinshaw:2012fq}
(temperature and polarization) with the routine for computing the
likelihood supplied by the WMAP team.  Then, we also add CMB data from
the SPT'11~\cite{Hou:2012xq,Story:2012wx}.  In order to address
foreground contributions, the Sunyaev--Zeldovich amplitude, $A_{\rm
  SZ}$, the amplitude of the clustered point source contribution,
$A_{\rm C}$, and the amplitude of the Poisson distributed point source
contribution, $A_{\rm P}$, are added as nuisance parameters in the CMB
SPT'11 data analysis.  Separately, we also consider the most recent
high multipole data from the ACT CMB experiment~\cite{Sievers:2013wk}
to explore the differences in the constraints on annihilating DM
scenarios arising when considering WMAP9 plus either SPT'11 or ACT10
data sets.  In addition to the CMB basic data sets, we add the latest
constraint on the Hubble constant $H_0$ from the
HST~\cite{Riess:2011yx} and galaxy clustering measurements, which are 
considered in our analyses via BAO signals.  We use here the BAO
signal from DR9~\cite{Anderson:2012sa} from data of the Baryon
Acoustic Spectroscopic Survey
(BOSS)~\cite{Schlegel:2009hj,Dawson:2012va}, with a median redshift of
$z=0.57$.  Together with the CMASS DR9 data, we also include the
recent measurement of the BAO scale based on a reanalysis (using
reconstruction~\cite{Eisenstein:2006nk}) of the Luminous Red Galaxies
(LRG) sample from Data Release 7 with a median redshift of
$z=0.35$~\cite{Padmanabhan:2012hf}, the measurement of the BAO signal
at a lower redshift $z=0.106$ from the 6dF Galaxy Survey
(6dFGS)~\cite{Beutler:2011hx} and the BAO measurements from the
WiggleZ Survey at $z=0.44$, $z=0.6$ and $z=0.73$~\cite{Blake:2011en}.

\subsection{Results}
\label{sec:results}

\begin{figure}[t]
\bc
\vspace{-1cm}
\includegraphics[width=0.8\textwidth]{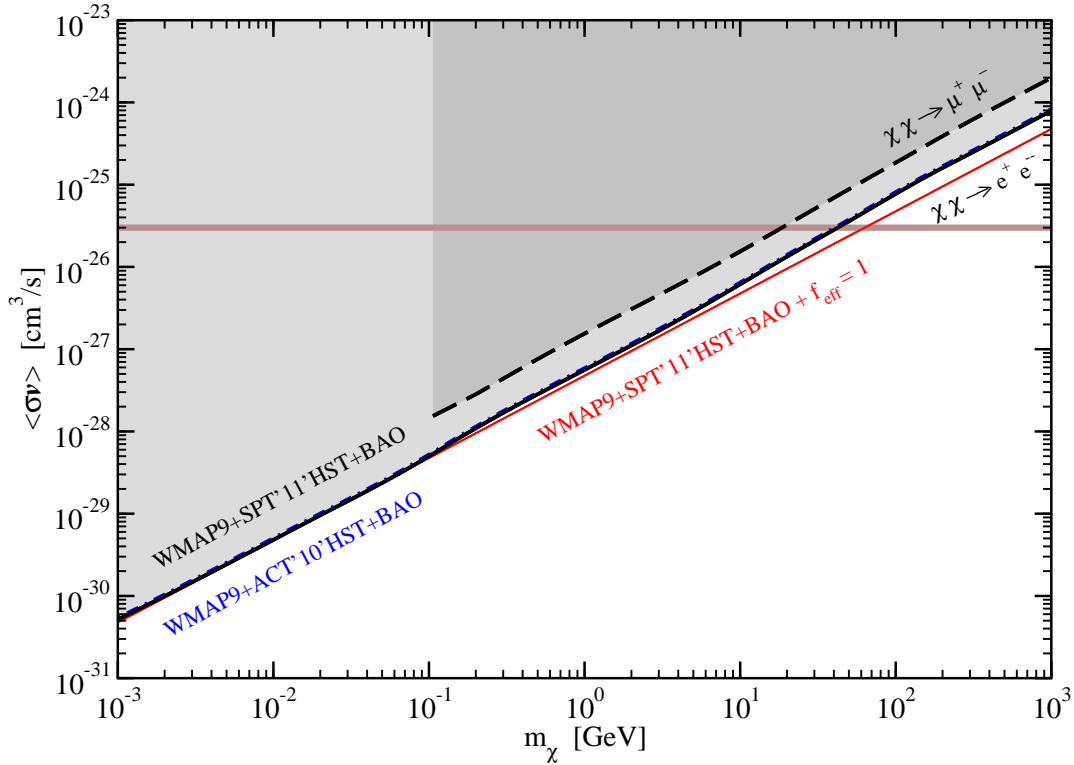} 
\ec
  \caption{\sl $95\%$~CL limits on the DM annihilation cross section
    $\sv$ as a function of the DM mass $\mx$ for the ``SPT''
    (WMAP9+SPT'11+HST+BAO) and ``ACT'' (WMAP9+ACT'10+HST+BAO)
    datasets.  The middle solid black (SPT'11) and dot--dashed blue
    (ACT'10) lines assume an effective deposition efficiency $\feff
    (\mx)$ for $\chi\chi \rightarrow e^+e^-$.  For the lower solid red
    line (SPT'11) we assume perfect efficiency, $f(z, \mx) = 1$, for
    $\chi\chi \rightarrow e^+e^-$.  The upper dashed black line
    (SPT'11) shows the bounds for $\chi \chi \rightarrow \mu^+\mu^-$
    with the corresponding effective energy deposition efficiency
    $\feff (\mx)$ for this channel.  Note that the results are
    identical when we include the contribution from DM annihilation in
    halos as parametrized in this work (see the text).  We also show
    the value of the canonical thermal annihilation cross section,
    $\sv = 3 \times 10^{-26} {\rm cm}^3 {\rm s}^{-1}$.}
  \label{fig:exclusion}
\end{figure}

Fig.~\ref{fig:exclusion} summarizes our main results in the
$\mx$--$\sv$ plane and shows the exclusion region extracted from the 
effect of this type of signatures on the CMB using the most recent
observations\footnote{Let us note that the best--fit values for the
  other six cosmological parameters fully agree with the results 
  without DM.  For instance, we obtain $z_{\rm reio} = 10.3 \pm 0.9$
  to be compared with $z_{\rm reio} = 10.6 \pm 1.1$, from
  WMAP9~\cite{Hinshaw:2012fq}.}.  The middle solid black line
represents the $95\%$ confidence level (CL) exclusion limit for the
case of DM annihilations into $e^+e^-$ from the smooth DM background,
when analyzing the SPT'11 data set plus WMAP9, BAO and HST, as described
above, and using the appropriate $\feff(\mx)$ efficiency function,
given in Eq.~(\ref{eq:feff}).  The dot--dashed blue line shows the
equivalent limit using ACT'10 data set plus WMAP9, BAO and HST.  The
lower solid red curve depicts the analogous $95\%$~CL exclusion limit
assuming the energy deposited equals the energy injected ($f(z, \mx) =
1$) using WMAP9+SPT'11+HST+BAO data sets.  Notice that, when including
the effective efficiency $\feff(\mx)$, the $\chi \chi \rightarrow e^+
e^-$ bounds become less stringent as the DM mass increases compared to
those when assuming perfect efficiency ($f(z, \mx) = 1$), as expected
from the left panel of Fig.~\ref{fig:feff_geff}.  Finally, the upper
dashed black line illustrates the $95\%$~CL bounds for the
$\mu^+\mu^-$ channel, with the corresponding $\feff(\mx)$ function, 
and using WMAP9+SPT'11+HST+BAO data sets.  In this case, the bounds are
weaker since $\sim 2/3$ of the energy is lost in the form of neutrinos. 

As a comparison with previous studies, we provide the limits on
$\pann$, which is often quoted in the literature, and is defined as:
\begin{equation}
\pann \equiv \feff(\mx) \, \frac{\sv}{\mx} ~.
\label{eq:pann}
\end{equation}
In Table ~\ref{table:pann} we present the bounds illustrated in
Fig.~\ref{fig:exclusion}, as well as our results for WMAP7 data with
previous releases of either ACT or SPT data in comparison with some
limits from the recent literature. In addition, we show the
improvements of individual data sets by adding only one or the other.
Tab.~\ref{table:pann} illustrates that while the improvements from the
inclusion of HST and BAO priors are marginal, the bounds improve
significantly by using more recent CMB data.  We attribute part of the
difference to the tighter error bars in the WMAP9 polarization data,
but find that the our improved bounds are mainly driven by the  better
accuracy at high $\ell$  of the recent ACT and SPT data releases. 

We find that the inclusion of annihilating DM in halos does not modify
the exclusion regions depicted in Fig.~\ref{fig:exclusion} for the
realistic halo model considered in this work and described in
Appendix~\ref{sec:halos}.  We have recomputed the $95\%$~CL upper
bounds, finding no improvement with respect to the bounds obtained
when only the contribution from the smooth DM component was considered.
The effects of the halo contribution could only be significant with an
enhancement of $g(z, \mx)$ of at least two orders of magnitude.  An
increase of about an order of magnitude could be obtained by using a
cuspier density profile for the DM halos than NFW.  In addition, a
decrease by four orders of magnitude in the uncertain and
model--dependent minimum halo mass ($M_{\rm min} = 10^{-10} M_\odot$)
would increase the maximum value of $g(z, \mx)$ only by a modest
factor of $\sim 2$.  Therefore, except from very extreme halo
models\footnote{Nevertheless, current N--body simulations can only
  resolve relatively massive halos at all redshifts (the limiting mass
  in the simulations we use is $\sim 10^5 M_\odot$ in
  Ref.~\cite{Prada:2011jf} and $\sim 10^{12} M_\odot$ at $z=0$ in
  Ref.~\cite{Watson:2012mt}) and hence these results rely on
  extrapolations to high redshifts and very low mass halos, which are
  the ones that are expected to contribute the most.  In this regard
  the contribution to the injected energy from DM halos is subject to
  important uncertainties.}, a large change in $g(z,\mx)$ so that the
halo contribution is relevant for CMB constraints, seems difficult to
be achieved. Finally, we point out that measurements of the IGM
temperature do not further constrain this model (see
Fig.~\ref{fig:Tm}) and hence the inclusion of a prior on the IGM
temperature would not modify the bounds summarized in
Fig.~\ref{fig:exclusion}.

\begin{table}[t]
\begin{center}
\begin{tabular}{c c c}
\hline \hline
Dataset & & $\pann$ [$10^{-6}$ m$^3$ s$^{-1}$ kg$^{-1}$] \\ \hline
WMAP7 + ACT'08  &(Galli \textit{et al.}~\cite{Galli:2011rz}) &
$<$ 1.17 \\ 
WMAP7 + SPT'09 & (Giesen \textit{et al.}~\cite{Giesen:2012rp}) &
$<$ 0.91 \\ \hline
*WMAP7 + SPT'09 & \multirow{2}{*}{this study} & $<$ 0.81 \\
WMAP7 + SPT'09 &						& $<$ 0.64 \\
WMAP9 + SPT'09 &  &$<$ 0.44 \\
WMAP9 \textit{only} &  &$<$ 0.66 \\
WMAP7 + SPT'11  &  &$<$ 0.32 \\
WMAP9 + SPT'11  & & $<$ 0.27 \\
WMAP9 + ACT'10  &  &$<$ 0.29 \\
\hline \hline
\end{tabular}
\end{center}
\caption{\sl Comparison between this study and previous results of
  $95\%$~CL limits on $\pann$, defined in Eq.~(\ref{eq:pann}). Our
  results, labeled ``this study'' also include HST and BAO data,
  which were found to have a very small impact on our $\pann$
  constraints. All results labeled ``this study'' were obtained using the priors in Tab. I, except the first line (labeled with an asterisk *) which used the optical depth $\tau$ rather than the reionization redshift $z_{\rm reio}$ with priors $\tau \in \{0.01, 0.8\}$. We include this result to facilitate comparison with other studies. }
\label{table:pann}
\end{table}

\section{Conclusions}

New CMB measurements have recently become available.  The WMAP team
has released their final nine--year data~\cite{Hinshaw:2012fq} and new
and precise measurements of the CMB damping tail from the
SPT~\cite{Hou:2012xq,Story:2012wx} and the ACT~\cite{Sievers:2013wk}
teams are also publicly available.  It is therefore timely to exploit
these new cosmological data sets to find bounds on different physics
models which may leave signatures on CMB observables.   

One of these scenarios is the injection of electromagnetic energy from
the annihilations of DM particles from the smooth DM background, which
would leave an imprint in both, the temperature and the polarization
CMB spectra.  After the recombination period, the residual ionization
fraction is increased, smearing the visibility function and inducing a
damping of the acoustic oscillations.  At late times, DM annihilation
in halos could also modify the reionization history of the Universe.
In this work we have addressed these two effects, by making use of the
numerical codes CAMB, CosmoRec and \texttt{cosmomc} to account for and
analyze the energy deposited into the CMB by DM annihilation products.
For that purpose, we have computed the energy deposition efficiency as
a function of the redshift and of the DM particle mass for the case of
$e^+ e^-$ and $\mu^+ \mu^-$ annihilation channels.

We have performed MCMC analyses to the most recent CMB measurements
(WMAP9 and separately adding SPT'11 and ACT'10 data sets), along with
BAO data and the HST prior on the Hubble constant $H_0$.  We have then
computed the mean values and errors of the six ``vanilla'' parameters
in Eq.~(\ref{eq:parameters}) plus the bounds on the DM annihilation
cross section $\langle \sigma v\rangle$ for DM masses ranging from
1~MeV to 1~TeV, which are shown in Fig.~\ref{fig:exclusion} and
represent our main result.  For instance, with the WMAP9, SPT'11, BAO 
and HST measurements, we find a $95\%$~CL upper bound of $\langle
\sigma v \rangle < 7.3\cdot 10^{-28}$~cm$^3$/s for the $e^+e^-$
channel for a DM mass of $\mx = 1$~GeV.  We exclude the thermal
annihilation cross section for masses below 30~GeV for a $e^+e^-$
final state and 15~GeV for the $\mu^+\mu^-$ final state.

Within a mixed reionization scenario, which includes both reionization
from first stars plus the contribution to the free electron fraction
from DM annihilations in halos (which have been modeled using recent
N--body simulations), the constraints from the smooth DM component
remain unchanged, even when large uncertainties on the halo model
parameters are addressed.  Upcoming data from the Planck telescope is
expected to improve the current CMB constraints on annihilating/decaying
scenarios and a detailed analysis will be carried out elsewhere.

\section*{Acknowledgments}

We thank C.~Evoli, S.~Pandolfi, F.~Prada, T.~Slatyer and W.~A.~Watson
for valuable discussions.  LLH is supported through an
``FWO--Vlaanderen'' post--doctoral fellowship project number 1271513.
LLH also recognizes partial support from the Belgian Federal Science
Policy Office through the Interuniversity Attraction Pole P7/37 and
from the Strategic Research Program ``High--Energy Physics'' of the
Vrije Universiteit Brussel.  OM is supported by the Consolider Ingenio
project CSD2007--00060, by PROMETEO/2009/116, by the Spanish Grant
FPA2011--29678 of the MINECO.  SPR is supported by the Spanish Grant
FPA2011--23596 of the MINECO and by the Portuguese FCT through
CERN/FP/123580/2011 and CFTP--FCT UNIT 777, which are partially funded 
through POCTI (FEDER).  ACV acknowledges support from FQRNT and
European contracts FP7--PEOPLE--2011--ITN.  OM and ACV are also
supported by PITN--GA--2011--289442--INVISIBLES.

\appendix
%%%%%%%%%%%%%%%%%%%%%%%%%%%%%
\section{Spectra of injected energy}
%%%%%%%%%%%%%%%%%%%%%%%%%%%%%
\label{sec:spectrum}

The simplest case is the direct annihilation to an $e^+e^-$ pair, in
which case the electron spectrum is a simple delta function: 
\begin{equation}
\frac{\ud N_{\rm e}}{\ud y_{\rm e}} = \delta(1 - y_{\rm e}) ~,
\end{equation}
where $y_{\rm e} = E_{\rm e}/\mx$. The electron/positron spectrum from
the $\chi \chi \rightarrow \mu^+\mu^-$ channel (in the limit $m_{\rm
  e} \ll m_\mu$) is given by~\cite{Mardon:2009rc} 
\begin{equation}
\frac{\ud N_{\rm e}}{\ud y_{\rm e}} = \frac{5}{3} - 3y_{\rm e}^2 +
\frac{4}{3}y_{\rm e}^3 ~.
\end{equation}
In either case, the dominant source of photons from this process comes
from internal bremsstrahlung\footnote{We have neglected the
  contribution from $\mu \rightarrow e \, \nu_e \, \nu_\mu
  \gamma$, which amounts to corrections to our final results below the
  percent level.}~\cite{Beacom:2004pe, Mardon:2009rc}
\begin{equation}
\frac{\ud N_\gamma}{\ud y_\gamma} = \frac{\alpha}{\pi}\frac{1+
  (1-y_\gamma)^2}{y_\gamma}\left(-1 + \ln
\left(\frac{4(1-y_\gamma)}{\epsilon_\ell}\right)\right) ,
\end{equation}
where $y_\gamma = E_\gamma/\mx$ and $\epsilon_\ell = m_\ell/m_\chi$
where $\ell = \{\rm e,\mu\}$, depending on the annihilation channel. 

Note that these expressions represent the spectra per outgoing particle
and must be multiplied by 2 to account for the two produced particles
per DM annihilation.

%%%%%%%%%%%%%%%%%%%%%%%%%%%%%%
\section{Contribution from DM halos to energy deposition}
\label{sec:halos}
%%%%%%%%%%%%%%%%%%%%%%%%%%%%%%%

The energy injected into the IGM by DM annihilation in halos at
redshift z is given by
\begin{equation}
\left(\frac{\ud E}{\ud V \ud t}\right)_{\rm halo, injected} = \zeta \,
\frac{\langle \sigma v \rangle}{m_\chi} \, \int \ud M
  \frac{\ud n_{\rm halo}}{\ud M}(M,z) \, \int_0^{r_{\Delta}} \ud r \, 4
\pi r^2 \, \rho_{\rm halo}^2(r) ~,
\label{eq:dEdVh}
\end{equation}
where the first integral represents the sum of the contributions from
all halos and the second integral is the contribution from a single
halo. 

Let us start by describing the contribution from a halo of mass $M$.  In
this term, $r_{\Delta}$ is the radius of a spherical halo at which
the mean matter density enclosed within is $\Delta$ times the critical
density of the Universe at redshift $z$, $\rho_{\rm c} (z) = \rho_{\rm
  c,0} \, (\Omega_{\rm m} \, (1+z)^3 + \Omega_\Lambda)$, with $
\rho_{\rm c,0}$ the critical density at $z=0$, so that the halo mass,
$M$, is
\begin{equation}
M = \Delta \, \rho_{\rm c} (z) \, \frac{4\pi}{3} \, r^3_{200} ~. 
\label{eq:delta200}
\end{equation}
The injection energy due to DM annihilations depends on the square of
the DM density profile.  In this work we consider spherical halos with
density profiles described by the NFW profile~\cite{Navarro:1995iw}, 
\begin{equation}
\rho_{\rm halo}(r) = \rho_{\rm s} \, \frac{4}{(r/r_{\rm s}) \,
  (1+r/r_{\rm s})^2} ~,
\label{eq:NFW}
\end{equation}
where $r_{\rm s}$ is the scale radius and $\rho_{\rm s}$ the density
at that radial distance.  Using this profile, the second integral in
Eq.~(\ref{eq:dEdVh}) can be computed analytically, 
\begin{eqnarray}
\int_0^{r_{\Delta}} dr \, 4 \pi r^2 \, \rho_{\rm halo}^2(r) & = &
\frac{4\pi}{3} \, (4 \rho_{\rm s})^2 \, r_{\rm s}^3 \,
\left(1-\frac{1}{(c_{\Delta}+1)^{3}}\right)\cr  & = & \tilde
g(c_{\Delta}) \, \frac{M \, \Delta \, \rho_{\rm c}(z)}{3}
\label{eq:shalo}
\end{eqnarray}
where we have used the concentration parameter $c_\Delta =
r_\Delta/r_{\rm s}$ and the halo mass $M$ as the parameters defining the
density profile, instead of $r_{\rm s}$ and $\rho_{\rm s}$.  The function
$\tilde g(c_\Delta)$ is given by 
\begin{equation}
 \tilde g(c_\Delta) = \frac{ c^3_\Delta \, \left[ 1 - (1+c_\Delta)^{-3}
     \right ] }{ 3 \left[ \ln (1+c_\Delta) - c_\Delta(1+c_\Delta)^{-1}
     \right ]^2} ~.  
\end{equation}

It is well known that halo concentrations depend on the halo mass and
redshift.  Hence, the two--parameter function describing the NFW
profile can be reduced to a one--parameter description for each given
redshift, and thus Eq.~(\ref{eq:shalo}) depends only on $M$ and $z$. 

A parametrization for $c_{200} (M,z)$ using $\sigma(M,z)$ was obtained
after a fit of all available data from the MultiDark/BigBolshoi
simulations\footnote{Note that these simulations have a limited range
  of validity in halo mass and redshift.  Outside this range, we
  extrapolate the fitting function and set a cutoff on the
  concentration parameter, $c_{200} = 100$, so that it does not
  diverge for high redshifts, although the exact value of this cutoff
  is not important.}~\cite{Prada:2011jf},  
\begin{equation}
  c_{200}(M,z) = B_0(x) \, \mathcal{C}(\sigma') ~, 
\label{eq:conc}
\end{equation}
where 
\begin{equation}
\mathcal{C}(\sigma')  =  A \,
\left[\left(\frac{\sigma'}{b}\right)^c+1\right]  
\exp\left(\frac{d}{\sigma'^2}\right) \hspace{1cm}; \hspace{0.5cm} 
A=2.881,\, b=1.257,\, c=1.022,\, d=0.060
\label{eq:csp} 
\end{equation}
and
\begin{equation}
\sigma' = B_1(x)\,
\sigma(M,x) \hspace{1cm}; \hspace{0.5cm} 
x \equiv \left( \frac{\Omega_{\Lambda,0}} {\Omega_{\rm m,0}}
\right)^{1/3} \frac{1}{1+z} ~, 
\label{eq:sp} 
\end{equation}
where $\Omega_{\rm m,0}$ is the matter density contribution at $z =
0$ and $\Omega_{\Lambda}$ the contribution to the density from the
cosmological constant.  The functions $B_0(x)$ and $B_1(x)$ are
defined as 
\begin{equation}
B_0(x) = \frac{c_{\rm min}(x)}{c_{\rm min}(1.393)} ~, \quad
B_1(x) = \frac{\sigma^{-1}_{\rm min}(x)}{\sigma^{-1}_{\rm
    min}(1.393)} ~, 
\label{eq:b0b1} 
\end{equation}
where 
\begin{eqnarray}
c_{\rm min}(x) & = & c_0 + (c_1- c_0)
    \left[\frac{1}{\pi}\arctan\left[\alpha(x-x_0)\right] + \frac{1}{2}
      \right]~, \label{eq:cmin} \\ 
\sigma^{-1}_{\rm min}(x) & = & \sigma^{-1}_0+
(\sigma^{-1}_1-\sigma^{-1}_0)
\left[\frac{1}{\pi}\arctan\left[\beta(x-x_1)\right] +\frac{1}{2} 
       \right] \label{eq:smin} 
\end{eqnarray}
with
\begin{equation}
c_0 = 3.681,\, c_1 = 5.033,\, \alpha = 6.948,\, x_0 = 0.424,
\label{eq:pars2}
\end{equation}
and
\begin{equation}
 \sigma^{-1}_0 = 1.047,\,  \sigma^{-1}_1 = 1.646,\, \beta = 7.386,\,
 x_1 = 0.526 ~. 
\label{eq:pars3}
\end{equation} 
The rms density fluctuation $\sigma(M, z)$ appearing in
Eq.~(\ref{eq:sp}) is defined as
\begin{equation}
\sigma^2(M, z) = \left(\frac{D(z)}{D(0)}\right)^2 \, \int \frac{dk}{k}
\, \frac{k^3 P(k)}{2 \pi^2} \, \left |\tilde W(k R) \right |^2 
\label{eq:sigmadef} 
\end{equation}
where $P(k)$ is the linear matter power spectrum and $W$ the Fourier
transform of the real--space top--hat window function of radius $R = (3
\, M /(4 \, \pi \, \rho_{\rm m,0}))^{1/3}$, which is given by $\tilde
W(kR) = \frac{3}{(kR)^3} \, \left[ \sin kR - (kR)   \cos kR \right ]$.
A good approximation for $\sigma(M,0)$ in the range $M\in[10^{-9},
  10^{17}]~{\rm M}_\odot$ is
\begin{equation}
\ln \sigma^{-1} (M,0) = 0.2506 \, \left(\frac{M}{\rm
  M_\odot}\right)^{0.07536} - 2.6 \, \left(\frac{M}{\rm
  M_\odot}\right)^{0.001745} ~.
\label{eq:sigmaappr}
\end{equation}

We have obtained $\sigma(M, 0)$ using the linear power spectrum $P(k)$
generated with CAMB package~\cite{Lewis:1999bs} assuming $\Omega_{\rm
  m} = 0.27, \Omega_\Lambda = 0.73, h = 0.7,\Omega_{\rm b} = 0.044$
and $n_{\rm s} = 0.96$, and then normalized to $\sigma_8 = 0.8$, which
were the parameters considered in the N-body simulations in
Ref.~\cite{Watson:2012mt} that we have used in our analysis.  We had 
to continue the spectrum generated by CAMB for $k>10^4$/Mpc.  For that
purpose,  we made use of a quadratic fit to $\log[P(\log(k))]$ and $k$
within the $[10,10^4]$/Mpc interval.

The growth factor is defined as
\begin{equation}
D(z)  =  \frac{5}{2}\left( \frac{\Omega_{\rm
    m,0}}{\Omega_{\Lambda,0}} \right)^{1/3} \frac{\sqrt{1+x^3}}{x^{3/2}}
    \int_0^x\frac{x^{3/2}dx}{[1+x^3]^{3/2}} ~,
\label{eq:dz}
\end{equation}
with $x$ given in Eq.~(\ref{eq:sp}), and it can also be approximated
as~\cite{Lahav:1991wc, Carroll:1991mt} 
\begin{equation}
D(z) = \left(\frac{(5/2) \, \Omega_{\rm m}(z)}{\Omega_{\rm
    m}^{4/7}(z)-\Omega_{\Lambda}(z)+\left(1+\Omega_{\rm m}(z)/2\right) \,
  \left(1+\Omega_{\Lambda}(z)/70\right)}\right) \, \frac{1}{1+z} ~,
\label{eq:Dappr}
\end{equation}
with $\Omega_{\rm m}(z) = \Omega_{\rm m, 0} (1+z)^3/(\Omega_{\rm m,
  0} (1+z)^3 + \Omega_{\Lambda})$ and $\Omega_{\Lambda}(z) = 1 -
\Omega_{\rm m}(z)$.

\begin{figure}
\includegraphics[width=0.6\textwidth]{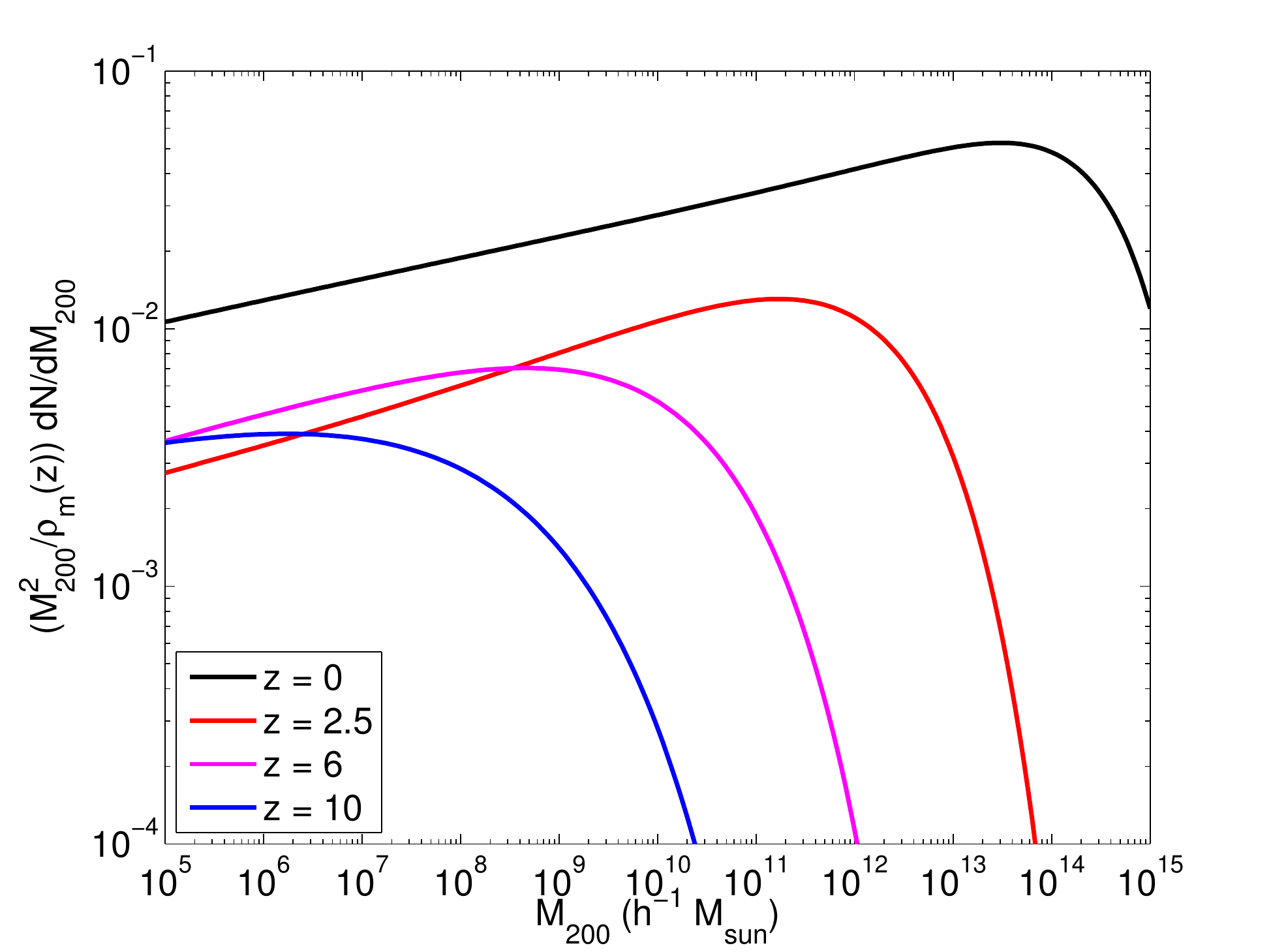}
\caption{\sl The halo mass function $\ud n/\ud M(M,z)$ from the
  N-body simulations from Ref.~\cite{Watson:2012mt} at various
  redshifts, increasing redshift from right to left.
\label{fig:halomass}}
\end{figure}

Once the contribution of a single halo is computed, one also needs the
halo mass function $\ud n_{\rm halo}/\ud M$ (Fig.~\ref{fig:halomass}),
the number of halos as a function of the halo mass and redshift, which
is parametrized as  
\begin{equation}
\frac{\ud n_{\rm halo}(M, z)}{\ud M} = \frac{\rho_{\rm m}(z)}{M^2} \,
\frac{\ud \ln \sigma^{-1}}{\ud \ln M} f(\sigma, z) ~,
\label{eq:dndM}
\end{equation}
where $\rho_{\rm m}(z) = \rho_{\rm m, 0} (1+z)^3$ is the average
matter density at redshift $z$ and $f(\sigma,z)$ is expected to be a
universal function\footnote{We keep the notation $f(\sigma)$ since it
  is widely used in the literature.  It should not be confused with
  $f(z)$, the energy deposition efficiency from the smooth DM
  component, which we use throughout the paper.} with respect to
redshift and changes in cosmology and can be parametrized
as~\cite{Tinker:2008ff}
\begin{equation}
f(\sigma, z) = A(z) \,
\left[\left(\frac{\sigma}{\beta(z)}\right)^{-\alpha(z)} + 1 \right] \,
e^{-\gamma/\sigma^2} 
\label{eq:fsigma} 
\end{equation}
where\footnote{This parametrization is based on the results obtained
  in Ref.~\cite{Watson:2012mt} by using the CubeP$^3$M halofinder
  (CPMSO)~\cite{HarnoisDeraps:2012vd} and the Amiga Halo Finder
  (AHF)~\cite{Gill:2004km, Knollmann:2009pb}.  In order to consider
  the contribution from all halos and at all redshifts, we have to use
  this fit outside the range where it was obtained, $-0.55 \leq
  \ln\sigma^{-1} \leq 1.35$ across all redshifts, which at $z=0$
  corresponds to halo masses between $2.6 \times 10^{12} \, {\rm
    M}_\odot$ and $10^{16} \, {\rm M}_\odot$.  This extrapolation is
  uncertain and there could be differences of up to a few orders of
  magnitude with respect to other parametrizations.  We have checked
  that the contribution from halos using the Press-Schechter
  formalism~\cite{Press:1973iz} with a critical linear overdensity
  for collapse $\delta_{\rm c} = 1.28$  as used in
  Refs.~\cite{Cirelli:2009bb, Giesen:2012rp} (instead of the
  conventional $\delta_{\rm c} = 1.686$, or $\delta_{\rm c} = 1.674$
  when considering the influence of dark energy~\cite{Lilje:1991ym,
    White:1992ri, Kochanek:1994vw, Eke:1996ds}) is 2--4 orders of
  magnitude larger than the parametrization we use, yet remains subdominant
  with respect to the smooth background contribution.}, for
$\Delta=178$,~\cite{Watson:2012mt}
\begin{eqnarray}
A_{178}(z) & = & \Omega_{\rm m}(z) \, \left(1.097 \, (1+z)^{-3.216} +
0.074\right) ~, \nonumber \\ 
\alpha_{178}(z) & = & \Omega_{\rm m}(z) \, \left(5.907 \, (1+z)^{-3.599} +
2.344\right) ~, \nonumber \\ 
\beta_{178}(z) & = & \Omega_{\rm m}(z) \, \left(3.136 \, (1+z)^{-3.068} +
2.349\right) ~, \nonumber \\ 
\gamma_{178} & = & 1.318 ~.
\label{eq:parsdndM}
\end{eqnarray}

In order to obtain the result for a different value of $\Delta$,
$f(\sigma, z)$ has to be scaled as~\cite{Watson:2012mt}
\begin{equation}
f_{\Delta}(\sigma, z) = \left[e^{(\frac{\Delta}{178}-1) \, (0.023 -
    0.072/\sigma^{2.13})} \, \left(\frac{\Delta}{178}\right)^{-0.456
    \, \Omega_{\rm m}(z)-0.139} \right] \, f_{178}(\sigma, z) 
\label{eq:fDelta}
\end{equation}

Hence, Eq.~(\ref{eq:dEdVh}) can be written as
\begin{eqnarray}
\left(\frac{dE}{dVdt}\right)_{\rm halo} & = & (1+z)^6 \, \zeta \,
\frac{\langle \sigma v \rangle}{m_\chi} \, \frac{\Delta \, \rho_{\rm
    m, 0}^2}{3 \, \Omega_{\rm m}(z)} \, \int_{M_{\rm min}}^{\infty}
\ud \log M \, 
\frac{\ud \ln \sigma^{-1}(M, z)}{\ud \log M} \, f_{\Delta}(\sigma
(M,z), z) \, \tilde g(c_\Delta (M,z))  ~.
\label{eq:haloinjectedEapp}
\end{eqnarray}
The function $G(z)$ defined in the text, Eq.~(\ref{eq:G}), is given by
\begin{equation}
G(z) = \left(\frac{\Omega_{\rm m, 0}}{\Omega_{\rm DM, 0}}\right)^2 \, \,
\frac{\Delta}{3 \, \Omega_{\rm m}(z)} \, 
\int_{M_{\rm min}}^{\infty}
\ud \log M \, 
\frac{\ud \ln \sigma^{-1}(M, z)}{\ud \log M} \, f_{\Delta}(\sigma
(M,z), z) \, \tilde g(c_\Delta (M,z))  ~,
\label{eq:Gapp}  
\end{equation}
and we take $\Delta = 200$. \\

Using this function, one can compute the efficiency $g(z, \mx)$,
Eq.~(\ref{eq:gofz}), that can be approximately given by
\begin{equation}
\geff(z, \mx) = \gamma(\mx) \, \Gamma(z)
\end{equation}
where
\begin{eqnarray}
\ln\gamma(\mx) & = & p_1 \, \ln(\mx/\textrm{GeV})^3 + p_2 \,
\ln(\mx/\textrm{GeV})^2 + p_3 \, \ln(\mx/\textrm{GeV}) + p_4~, \\ 
\ln\Gamma(z) & = & q_1 (1+z)^{q_2} + q_3 (1+z)^{q_4}
\label{eq:gfit}
\end{eqnarray}
with 
\begin{equation}
q_1  = -0.008262, \, q_2 =1.914, \, q_3 = 3.604, \, q_4 = -0.06479~.
\label{eq:gparam}
\end{equation}
The $p_i$ coefficients for the electron channel are:
\begin{equation}
p_1 = 0.006803, \,  p_2 = -0.05783, \, p_3 =  -0.8043, \, p_4 = 0.9186~,
\end{equation}
and for the muon channel:
\begin{equation}
p_1 = 0.004854, \, p_2 = -0.05132, \, p_3 = -0.7153, \, p_4 = 0.5491~,
\end{equation}
These approximations do not induce errors larger than a few percent
with respect to the case of using the full $g(z, \mx)$.

\bibliographystyle{utphys}
\bibliography{efficiency.bib}

\end{document}